\documentclass[a4paper,aps,prb,twocolumn,a4paper]{revtex4}

\usepackage{epsf,subfigure}
\usepackage{amsmath,graphicx}
\usepackage[utf8]{inputenc}
\usepackage[T1]{fontenc}
\usepackage{times,ulem,color,bm}
\usepackage[squaren,Gray]{SIunits}

\newcommand{\cora}[1]{\textcolor{black}{#1}}

\usepackage[pdfstartview=FitV,bookmarks=true,
            linktocpage=true,colorlinks=true,
            urlcolor=blue,linkcolor=blue,
            citecolor=blue]{hyperref}

\begin{document}

\title{Negative reflection of elastic guided waves in chaotic and random scattering media}

\author{Benoît Gérardin}

\author{Jérôme Laurent}
\author{François Legrand}
\author{Claire Prada}
\author{Alexandre Aubry} \email{alexandre.aubry@espci.fr}
\affiliation{ESPCI Paris, PSL Research University, CNRS, Univ Paris Diderot, Sorbonne Paris Cité, Institut Langevin, UMR 7587, 1 rue Jussieu, F-75005 Paris, France}

\date{\today}

\begin{abstract}
The propagation of waves in complex media can be harnessed either by taming the incident wave-field impinging on the medium or by forcing waves along desired paths through its careful design. These two alternative strategies have given rise to fascinating concepts such as time reversal or negative refraction. Here, we show how these two processes are intimately linked through the negative reflection phenomenon. A negative reflecting mirror converts a wave of positive phase velocity into its negative counterpart and vice versa. In this article, we experimentally demonstrate this phenomenon with elastic waves in a 2D billiard and in a disordered plate by means of laser interferometry. Despite the complexity of such configurations, the negatively reflected wave field focuses back towards the initial source location, thereby mimicking a phase conjugation operation while being a fully passive process. The super-focusing capability of negative reflection is also highlighted in a monochromatic regime. The negative reflection phenomenon is not restricted to guided elastic waves since it can occur in zero-gap systems such as photonic crystals, chiral metamaterials or graphene. Negative reflection can thus become a tool of choice for the control of waves in all fields of wave physics.  
\end{abstract}

\maketitle

\section*{Introduction}

Controlling the propagation of acoustic or electromagnetic waves, is of fundamental interest for many applications ranging from imaging the living and detecting hazardous components, to information processing and structural health monitoring. In the past decades, there has been many proposals in this regards, which can be separated within two approaches. On the one hand, the wave fields can be tamed in order to take advantage of the complexity of propagation media, for instance, to focus waves or image various objects. This is realized in the temporal domain using time reversal (TR) mirrors~\cite{Fink,Lerosey,Fort} or in the spatial domain using phase conjugation (PC) \cite{Hellwarth:77,yariv,fischer} and wave-front shaping techniques~\cite{mosk2012controlling} developed in optics. On the other hand, one can force waves along desired paths through a careful design of man-made materials. This can be achieved using metamaterials, an arrangement of tailored sub-wavelength building blocks from which the material gains its unusual properties~\cite{pendry1999magnetism,Smith,Liu1734}. The advent of such structures has given rise to fascinating concepts such as negative refraction~\cite{veselago1968electrodynamics,pendry2000negative,shelby2001experimental}, transformation optics~\cite{pendry,leonhardt} or metasurfaces~\cite{capasso}. Although the concepts of TR and negative refraction have been developed in an independent fashion, they are intimately linked processes \cite{Maslovski,pendry2008time}. 

\begin{figure}[!ht]
\centering
\includegraphics[width=8cm]{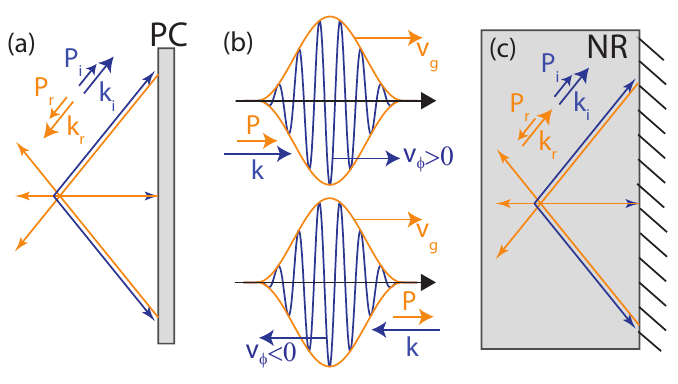}
\caption{Analogy between phase conjugation and negative reflection. (a) A PC mirror has the property to reverse both the propagation direction and the phase of an incident wave field. (b) A wave packet exhibiting a positive/negative phase velocity, $v_{\phi}$, is associated with parallel/antiparallel wave and Poynting vectors, respectively. (c) A NR mirror converts an incident wave of phase velocity $v_{\phi}$ into a wave of opposite phase velocity $-v_{\phi}$.}
\label{fig00}
\end{figure}

Here, we want to push forward this analogy by investigating the negative reflection (NR) phenomenon. A NR mirror is an interface at which light or sound is retro-reflected. There is a strong similarity with a PC mirror. In a PC experiment, if the incident wave is divergent, the PC wave is converging [Fig.~\ref{fig00}(a)]. It follows the same path as the incident wave but in an opposite way, thus back-focusing exactly on the source location. Let us now consider a NR mirror and an incident forward wave with a Poynting vector, $\mathbf{P_i}$, and a wave vector, $\mathbf{k_i}$, pointing in the same direction (\textit{i.e.} with a positive phase velocity, $v_{\phi}>0$) [see Fig.~\ref{fig00}(b,c)]. The NR mirror gives rise to a reflected backward wave ($v_{\phi}<0$), \textit{i.e.} with a Poynting vector, $\mathbf{P_r}$, and a wave vector, $\mathbf{k_r}$, of opposite direction [see Fig.~\ref{fig00}(b,c)]. By virtue of the Snell-Descartes law, the incident and reflected wave vectors are strictly identical ($\mathbf{k_r}=\mathbf{k_i}$) but their Poynting vectors shall be in opposite directions ($\mathbf{P_r}=-\mathbf{P_i}$) [Fig.~\ref{fig00}(b)]. Negative reflection is a perfectly reciprocal phenomenon: A backward incident wave with anti-parallel $\mathbf{k_i}$ and $\mathbf{P_i}$ will be negatively reflected into a forward mode with parallel $\mathbf{k_r}$ and $\mathbf{P_r}$.

If the incident wave is divergent, the NR wave back-converges towards the initial source position. We thus recover a similar effect as in PC, although the reflected wave vectors in each case are in opposite directions [Fig.~\ref{fig00}(a,c)]. The PC concept and its temporal equivalent, TR, have shown to be particularly powerful in reverberating or through inhomogeneous media \cite{Fink,derode,Draeger}. If the phase accumulated by the waves following each scattering path is reversed, the PC waves re-accumulate the same phase during their trip back due to spatial reciprocity. They finally sum up coherently at the initial source location. This has led to spectacular focusing experiments in acoustics and optics\cite{mosk2012controlling}. Any scattering or reverberating medium can become a lens if the incident wave is properly designed by a PC process. In this paper, we show how NR can provide similar focusing abilities in complex environments.  

NR may occur in a medium that can simultaneously support forward and backward waves. {Such a coexistence is possible in, at least, two types of physical systems. The first situation corresponds to elastic guided waves propagating, for instance, in isotropic~\cite{tolstoy1957wave,meitzler1965backward,mindlin1960waves,negishi1987existence} or anisotropic~\cite{poncelet} plates, pipes~\cite{meitzler1965backward}, rods~\cite{laurent}, strips~\cite{cross}, bi-layer structures~\cite{maznev}, or even in the Earth's crust \cite{geofisica}. Such systems can actually support an ensemble of modes, the so-called Lamb waves, which exhibit complex dispersion properties~\cite{royer}. In particular, some of these Lamb waves display a zero-group velocity (ZGV) point which results from the interference between forward and backward modes~\cite{poncelet,meitzler1965backward,tolstoy1957wave,mindlin1960waves,negishi1987existence,prada2008}. The second situation corresponds to zero-gap systems, which actually support two bands that cross symmetrically, without forming a band gap. One of the most notable zero-gap material is graphene~\cite{Cheianov}, or its photonic/phononic crystal analogues~\cite{joannopoulos}. Others are quarter wave stack photonic crystals~\cite{Sivan}, chiral metamaterials~\cite{pendry_chiral,Zhang} or transmission-line devices~\cite{Antoniades}. NR is therefore a very general phenomenon and this work can be, in principle, transposed to all the aforementioned systems. In this article, we have chosen to consider the NR of elastic waves in plates because they provide an excellent platform to investigate negative refraction~\cite{bramhavar2011negative,philippe2015focusing,legrand} and reflection~\cite{germano2012anomalous,veres2016broad,gerardin2016}. Indeed, laser interferometry allows to probe the wave-field over the whole plate surface with an excellent spatial, temporal and spectral resolution. The NR phenomenon and its analogy with PC will be investigated in two complex geometries: a chaotic billiard and a disordered plate. Whereas the reflection of waves at the boundaries of a cavity or by a random distribution of scatterers usually induces a fully random speckle wave field, NR will be shown to give rise to a purely deterministic wave field: the PC replica of the incident wave. At last, we will discuss about the manifestation of NR in the monochromatic regime. In particular, we will describe how it can lead to the super-focusing of elastic waves in a cavity or to wave trapping in a scattering medium.}

\section*{Results}

\subsection*{Forward and backward Lamb modes}

\begin{figure*}[!ht]
\centering
\includegraphics[width=14cm]{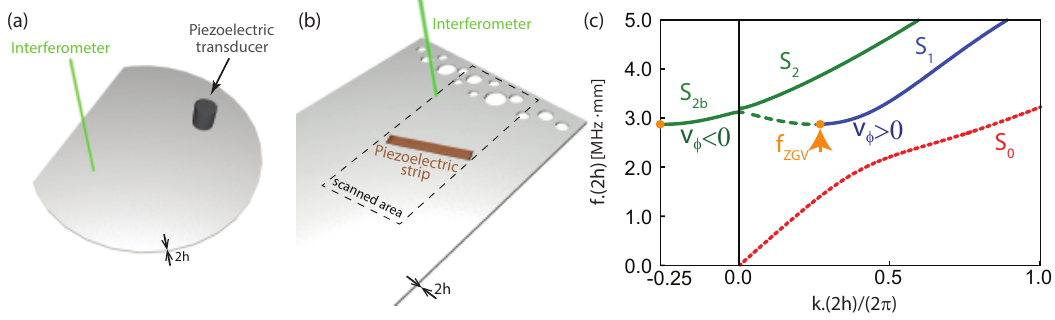}
\caption{Experimental configuration. (a) Lamb modes are generated in a Duralumin plate of thickness $2h = 1.5$ mm by a 7 mm-diameter transducer. The Duralumin plate is here a 80 mm-diameter truncated chaotic billiard. (b) Lamb modes are excited in a disordered Duralumin plate of thickness $2h = 2$ mm by a piezoelectric strip. In each case (a)-(b), the normal component of the plate vibration is measured with an interferometric optical probe, mounted on a 2D translation stage. (c) Dispersion curves of the first propagating symmetrical Lamb modes in a Duralumin plate computed from the Rayleigh-Lamb equation \cite{royer}.}
\label{fig1} 
\end{figure*}

Whereas an infinite medium supports two longitudinal and transverse modes traveling at unique velocities, plates support two infinite sets of Lamb wave modes. Lamb waves are elastic waves whose particle motion lies in the sagittal plane that contains the direction of wave propagation and the plate normal. The deformation induced by these modes can be either symmetric {(referred to as $S_i$)} or antisymmetric {(referred to as $A_i$)} with respect to the median plane. {All symmetric and antisymmetric Lamb modes (except the zero-order $S_0$, $A_0$, and first-order $A_1$ modes) can exhibit a ZGV point}~\cite{prada2008}. For the sake of simplicity, but without loss of generality, we will specifically focus on the first symmetric modes. 

In this article, we study the propagation of elastic waves across a Duralumin plate (aluminium alloy) of density \cora{$d = 2790~\textrm{kg/m}^3$}, longitudinal wave velocity $c_L = 6400~\textrm{m}.\textrm{s}^{-1}$, and transverse wave velocity $c_T = 3120~\textrm{m}.\textrm{s}^{-1}$. {The dispersion curves of Lamb modes in this material are displayed in Fig.~\ref{fig1}(c). The symmetric zero-order mode $S_0$ is the extensional mode of the plate. It exhibits free propagation to zero frequency, whereas the higher order modes admit a cut-off frequency. The dispersion branch $S_1$ displays several crucial features. As illustrated by Fig.~\ref{fig1}(c), its repulsion with the $S_{2}$-mode gives rise to a ZGV point $(f_{ZGV},k_{ZGV})$~\cite{prada2005laser,holland2003air}. In the case of Duralumin, $f_{ZGV} \times 2h=2.86$ MHz.mm and $k_{ZGV} \times 2h/(2\pi)=0.26$, with $2h$ the plate thickness. From the ZGV resonance frequency $f_{ZGV}$ to the cut-off frequency, there is a coexistence between a backward mode, referred to as $S_{2b}$ in the literature, and the forward mode $S_{1}$~\cite{meitzler1965backward}. 

In general, at a plate edge, a single Lamb mode is reflected into a large number of modes that can be either propagative, inhomogeneous or evanescent. The key phenomenon occurring with the $S_1$ Lamb mode in the vicinity of the ZGV point is that it is mainly reflected into the $S_{2b}$ mode. An analytical study of this phenomenon\cite{gerardin2016} has shown that the reflection coefficient of the $S_1$ mode into the $S_{2b}$ mode is close to $-1$, while the conversion into the other modes can be neglected. Furthermore, this peculiar property holds over a large angular spectrum close to the ZGV-point. This phenomenon arises from the equality of $S_1$ and $S_{2b}$ wave numbers at the ZGV-point. This implies that these two modes are associated with similar stress-displacement fields and only differ by their opposite Poynting vectors. As a consequence, the stress-free boundary condition at the edge of the plate can be satisfied with a simple combination of $S_1$ and $S_{2b}$ modes stress field, leading to a reflection coefficient $\rho = -1$. As a result, a free edge acts as a nearly perfect NR mirror for the $S_1$ and $S_{2b}$ modes~\cite{gerardin2016}. In this paper, we will now show how this peculiar property holds in much more complex geometries and how negative reflection can be taken advantage of to harness waves in chaotic or scattering media. 


\subsection*{Negative reflection in chaotic cavity}

This striking effect is first investigated in an elastic chaotic billiard [see Fig.~\ref{fig1}(a)]: a truncated circular plate of diameter 80 mm and thickness $2h=1.5$ mm. The ZGV-point occurs at $f_{ZGV}=1.91$ MHz and $k_{ZGV}=1.06$ mm$^{-1}$. The ultrasound source consists in a $7$~mm-diameter piezoelectric transducer placed on top of the plate. The transducer’s width is larger than the ZGV wavelength ($\lambda_{ZGV} \sim 6$ mm) so that a dominant excitation of the $S_{2b}$-mode is provided. The out-of-plane component of the induced wave-field is recorded in the time domain over 1 ms by means of laser interferometry (see Methods). A discrete Fourier transform of the recorded signals is first performed over the temporal domain. The $S_{2b}$ and $S_{1}$ contributions are then separated with low and high pass spatial filters, respectively, with a cut-off at $k_{ZGV}$ (see Supplementary Information S1). Figure \ref{fig2}(a) shows the incident $S_{2b}$ wave field in the vicinity of the ZGV frequency ($f=1.91$ MHz). Not surprisingly, it displays divergent cylindrical wave fronts centered around the transducer location.
\begin{figure*}[!ht]
\centering
\includegraphics[width=12cm]{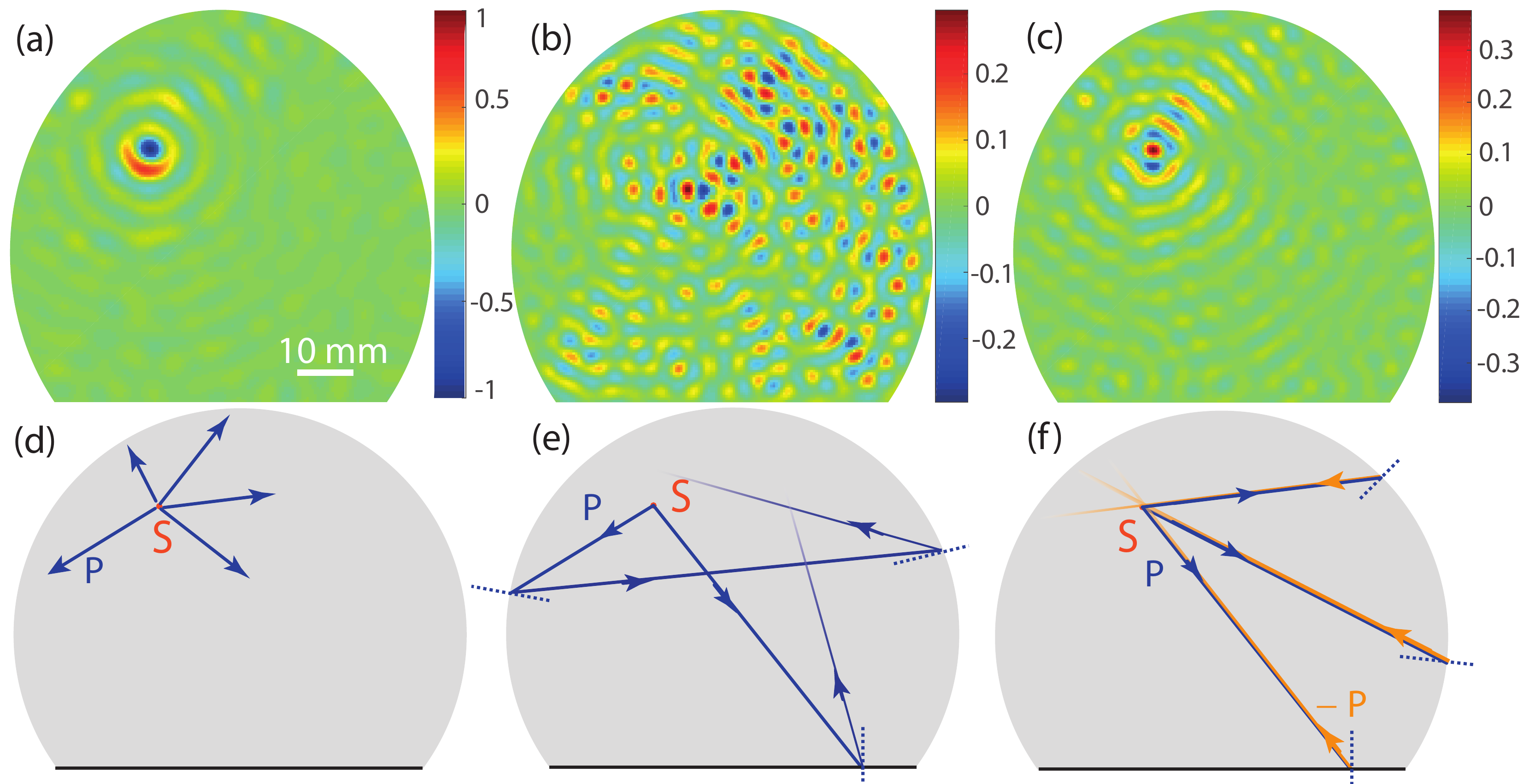}
\caption{Negative reflection in the truncated billiard. (a) Incident wave field associated with the excited $S_{2b}$-mode at frequency $f= 1.91$ MHz.  (b,c) Reflected wave field associated with the $S_{1}$-mode at frequency $f= 1.92$ MHz and $f= 1.91$ MHz, respectively. (d,e,f) Ray path trajectories involved in (a,b,c), respectively.}
\label{fig2}
\end{figure*}
Figures \ref{fig2}(b) and (c) show the reflected $S_1$ wave field at $f=1.92$ MHz and $f=1.91$ MHz, respectively. The first frequency is well above the ZGV-point displayed in Fig.~\ref{fig1}(c). There is a clear mismatch between the $S_1$ and $S_{2b}$ wave numbers: $k_{1}=1.26$ mm$^{-1}$ and $k_{2b}=0.86$ mm$^{-1}$, respectively (see Supplementary Information S1). As dictated by Snell's law, the reflection angle is, in this case, negative but not equal, in absolute value, to the incident angle [Fig.~\ref{fig2}(e)]. This explains the random feature displayed by the reflected wave field in Fig.~\ref{fig2}(b), which is characteristic of a chaotic cavity. In the vicinity of the ZGV frequency ($f=1.91$ MHz), the wave numbers of $S_1$ and $S_{2b}$ almost coincide: $k_{1}=1.145$ mm$^{-1}$ and $k_{2b}=0.975$ mm$^{-1}$ (see Supplementary Information S1). As a consequence, NR gives rise to a cylindrical wave front that converges back towards the initial source location [Figs.\ref{fig2}(c,f)], thus mimicking a PC experiment. Note that the NR wave field [Fig.~\ref{fig2}(e)] is out-of-phase compared to the incident wave front [Fig.~\ref{fig2}(a)], thereby confirming the theoretical prediction of a reflection coefficient, $\rho=-1$, between the $S_{1}$ and $S_{2b}$ modes at the ZGV-point \cite{gerardin2016}. As for a TR experiment in a cavity, the size of the focal spot originating from the NR process is limited to a half-wavelength~\cite{Draeger}. This can be interpreted in terms of the diffraction limit and the loss of the evanescent components~\cite{cassereau,rosny,carminati}.

The dynamics of the NR process is revealed by performing a discrete Fourier transform of the recorded signals over apodized time windows.
\begin{figure*}[!ht]
\centering
\includegraphics[width=12cm]{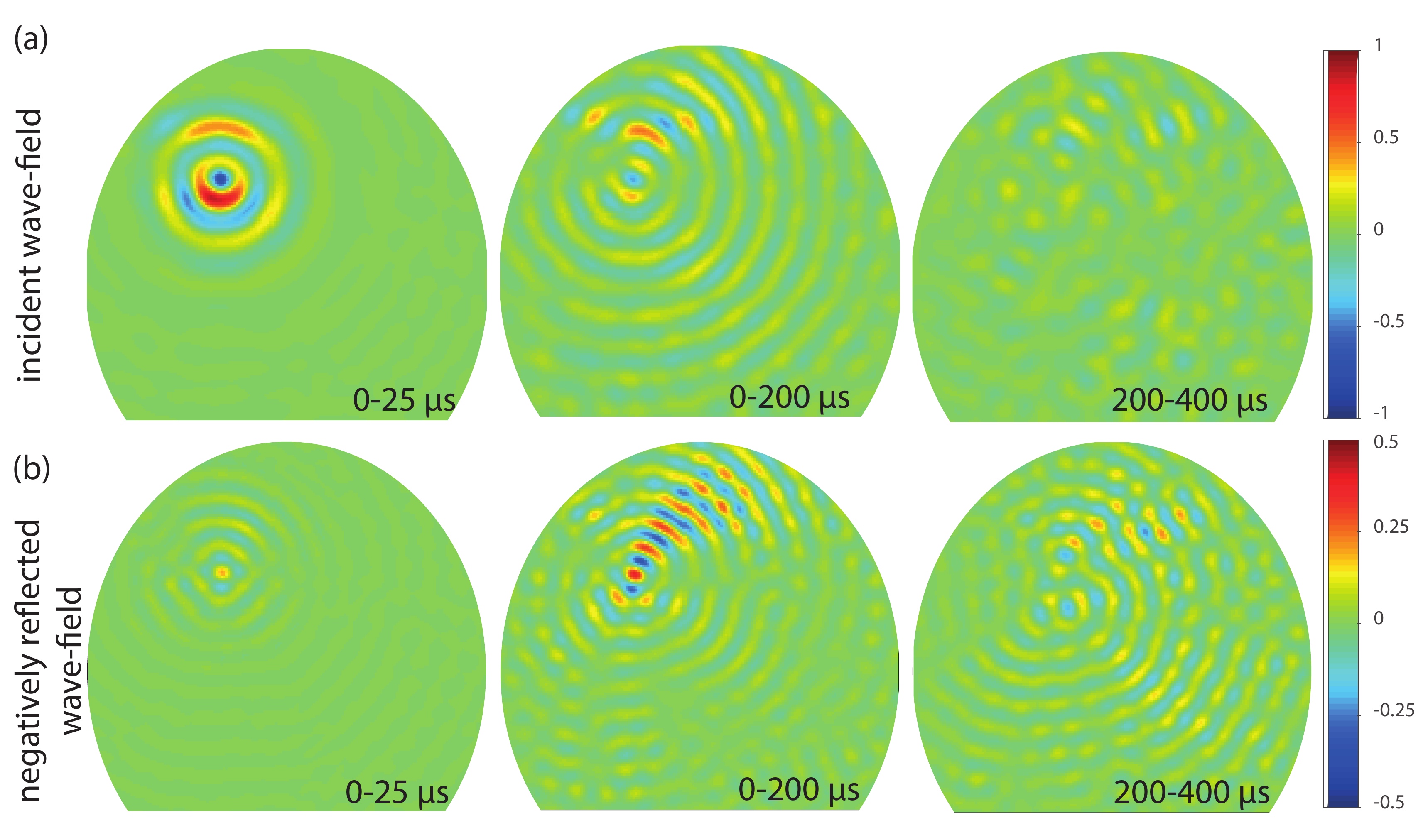}
\caption{Time evolution of the negative reflection process taking place in the truncated billiard. The $S_{2b}$ and $S_{1}$ wave fields are displayed in panels (a) and (b), respectively, over the indicated time windows. Note the different color scales in (a) and (b).}
\label{fig3}
\end{figure*}
The corresponding wave-fields are displayed in Fig.~\ref{fig3}. The top and bottom panels show the temporal evolution of the $S_{2b}$ and $S_{1}$ modes, respectively, at the central frequency $f =1.91$ MHz. In the time window $[0-25]$ $\mu$s (short time), the $S_{2b}$-mode is largely predominant confirming its selective excitation by the transducer. The incident wave front then propagates across the cavity and finally vanishes when it reaches the cavity boundary. The residual $S_{2b}$ random wave field observed at long times of flight ($[200-400]$ $\mu$s) can be explained by the spurious reflections undergone by the NR wave field when it converges back to the transducer. Fig.~\ref{fig3}(b) displays the time evolution of the $S_1$-mode. As soon as the incident $S_{2b}$-mode reaches the cavity boundary, it is converted into the NR $S_1$-mode. As in a PC or TR experiment, the NR wave field focuses back on the source location. A nice focus is obtained for instance over the time range $[0-200]$ $\mu$s. Beyond that time, we observe a progressive degradation of the focal spot. Longer NR paths are actually more sensitive on the slight mismatch between the wave numbers $k_1$ and $k_{2b}$ in the vicinity of the ZGV-point (see Supplementary Information S1). Figures \ref{fig2} and \ref{fig3} illustrate the remarkable properties of the NR phenomenon in a complex medium. Whatever the chaotic feature of the environment, the incident wave field is re-created, but traveling backward, retracing its passage through the medium and focusing back to the source location. However, unlike TR or PC, a NR mirror is fully passive. No energy is injected into the system through a TR mirror or a nonlinear wave mixing process.  

\subsection*{Negative reflection by a random disordered slab}
\begin{figure*}[!ht]
\centering
\includegraphics[width=16cm]{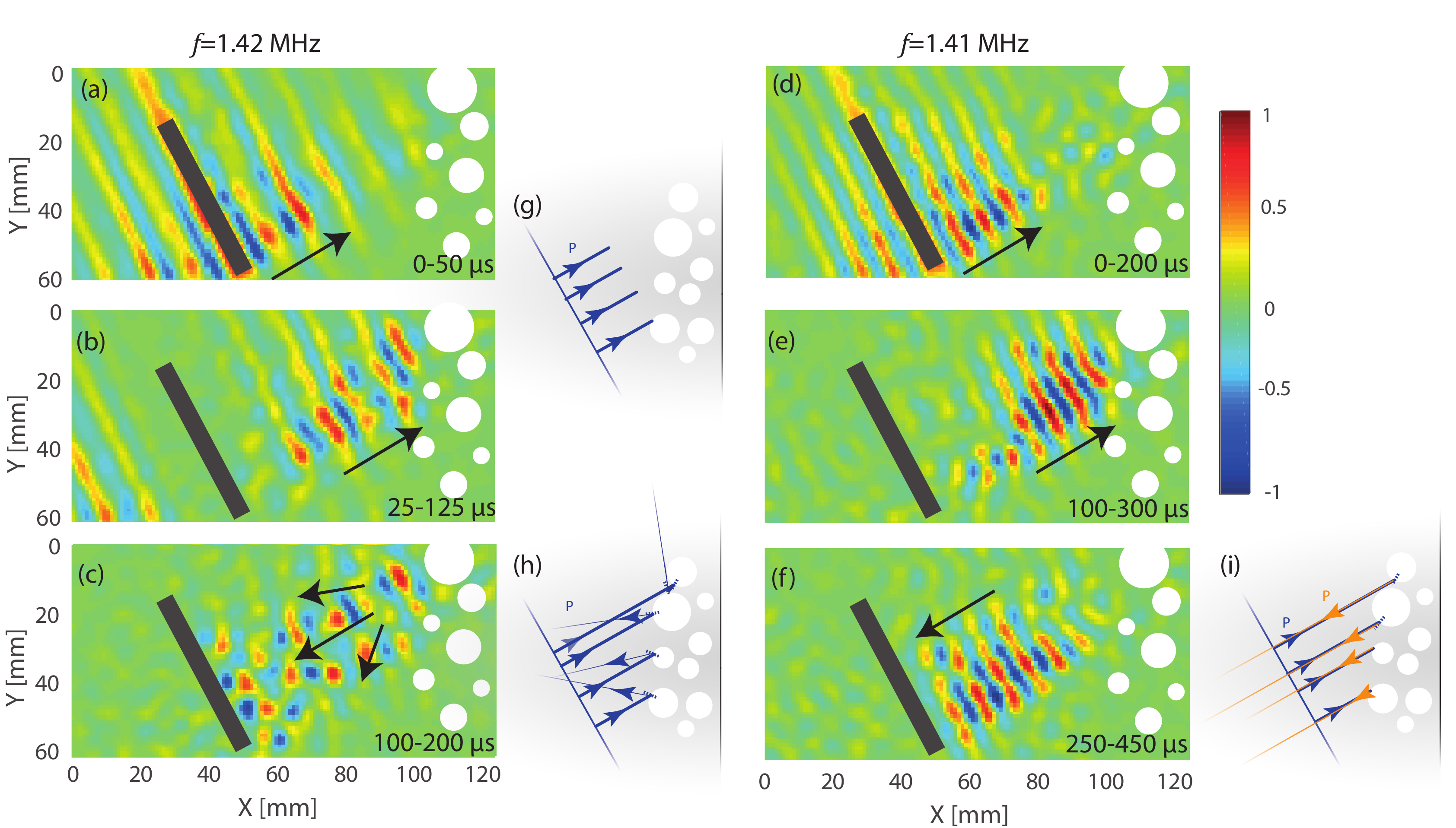}
\caption{Negative reflection at a disordered plate edge. Normal displacement field filtered at $1.42$ MHz (left) and $1.41$ MHz close to the ZGV point (right). The Fourier analysis of the recorded signals performed on the indicated time windows yields the incident (a,b,d,e) and reflected (c,f) wave-fields. Time windows are different to account for the frequency-dependence of the group velocity for Lamb modes. Ray path trajectories are illustrated for incident (g) and negatively reflected waves with $k_1 \neq k_{2b}$ (h) and $k_1=k_{2b}$ (i).}
\label{fig4}
\end{figure*}
Following the demonstration of the NR phenomenon in a cavity, we now address the more challenging case of a scattering layer made of randomly distributed holes. Such finite-size scatterers actually correspond to the small-scale limit of a curved plate edge. As NR occurs up to large angles of incidence~\cite{gerardin2016}, holes drilled in a plate constitute good candidates for finite-size NR mirrors.

The experiment is depicted in Fig.~\ref{fig1}(b). The system under study is a plate of thickness $2h \sim 2$ mm for which the ZGV-point occurs at $f_{ZGV}=1.41$ MHz and $k_{ZGV}=0.89$ mm$^{-1}$. Disorder is introduced by drilling circular holes with diameters ranging from 4 to 12.5 mm over a thickness of 20 mm from one of the free edge of the plate. The holes diameters are comparable with the wavelength $\lambda_{ZGV} \sim 8$ mm. The ultrasound source is a piezoelectric strip of 25 mm-length, 5 mm-width and 1 mm-thickness glued on top of the plate [Fig.~\ref{fig1}(b)]. This strip is located at 40 mm from the free edge and tilted 25$^o$ with respect to it. The source width ($\sim \lambda_{ZGV}/2$) allows to excite both the $S_1$ and $S_{2b}$ Lamb modes in the vicinity of the ZGV-point. The out-of-plane component of the local vibration of the plate is measured in the time domain during 1 ms over an area of dimension 60$\times$120 mm$^2$ (see Methods). In order to isolate the incident and reflected wave fronts in the vicinity of the ZGV-point, a discrete Fourier transform of the recorded signals is performed over distinct time windows. Figure \ref{fig4}(a)-(b) show the incident wave front at frequency $f=1.42$ MHz ($f \neq f_{ZGV}$), in the time windows $[0-50]$ $\mu$s and $[25-125]$ $\mu$s, respectively. Figure \ref{fig4}(d)-(e) displays the same wave-front but at frequency 1.41 MHz ($f \simeq f_{ZGV}$), for the time windows $[0-200]$ $\mu$s} and $[100-300]$ $\mu$s, respectively. The temporal windows considered at these two frequencies are different because the group velocity of Lamb modes strongly varies in the vicinity of the ZGV point. In both cases, the incident wave front is a wave packet of well-resolved momentum impinging the diffusive layer with an angle of incidence equal to 25$^o$ [Fig.~\ref{fig4}(g)]. This wave packet is then scattered by the holes [Fig.~\ref{fig4}(c,f)]. Figure \ref{fig4}(c) shows the reflected wave field in the time window $[100-200]$ $\mu$s and at frequency $f=1.42$ MHz, that is well above the ZGV-point displayed in Fig.~\ref{fig1}(c). The wave number mismatch between the $S_1$ and $S_{2b}$ modes results in a randomization of the reflected wave vector induced by the disordered layer [Fig.~\ref{fig4}(h)] and a speckle-like feature of the reflected wave field in Fig.~\ref{fig4}(c). On the contrary, in the vicinity of the ZGV frequency ($f=1.41$ MHz), the incident and reflected wave vectors point exactly in opposite directions [Fig.~\ref{fig4}(i)]. Figure \ref{fig4}(f) shows the corresponding NR wave field in the time window $[250-450]$ $\mu$s and at frequency $f=1.41$ MHz. Despite disorder, NR gives rise to a well resolved wave packet with a momentum opposed to the incident one. The scattering layer not only retro-reflects the incident wave packet but also preserves its amplitude and phase. To some extent, the scattering layer is thus cloaked by the NR phenomenon. Whatever the position and concentration of scatterers, the NR wave-field always displays the same feature and corresponds to the PC replica of the incident wave-field. This experiment thus illustrates the robustness of the NR phenomenon.

\section*{Discussion}
 
In this paper, the NR phenomenon is experimentally investigated in the time domain since it allows a clear discrimination between the incident and NR wave-fields. A first point we would like to discuss is the manifestation of the NR phenomenon in the frequency domain. NR can actually induce striking features in this regime. For sake of simplicity and generality, a scalar model is now considered to discuss about the potential impact of NR. Of course, it does not account for the full elastic problem but our aim is here to grasp the main features of NR for all kind of waves. Let us assume a monochromatic point source in a chaotic cavity [Fig.~\ref{fig1}(a)]. The corresponding diverging incident wave-field is given by the free space Green's function $G_0(\mathbf{r}|\mathbf{r_0})$. As shown in the Supplementary Information S2, multiple NR events on the cavity boundary have to be taken into account in the monochromatic regime. It yields the following expression for the Green's function $G(\mathbf{r}|\mathbf{r_0})$ in the cavity,
\begin{eqnarray}
\label{NR}
G(\mathbf{r}|\mathbf{r_0}) & = & \underbrace{G_0(\mathbf{r}|\mathbf{r_0})}_{\mbox{incident wave-field}} \nonumber \\
& + & \underbrace{\frac{\rho}{1-\rho} \left [ G_0(\mathbf{r}|\mathbf{r_0}) - G_0^*(\mathbf{r}|\mathbf{r_0}) \right ]}_{\mbox{NR wave-field}}.
\end{eqnarray} 
where the symbol $^*$ stands for phase conjugate. The NR wave-field is made of two contributions. The first component is a wave converging towards the initial source at $\mathbf{r_0}$, of opposite phase velocity compared to the incident wave-field. This wave is accounted for by $G_0(\mathbf{r}|\mathbf{r_0})$ in the second term of Eq.~(\ref{NR}). Owing to energy flux conservation~\cite{rosny}, this wave does not stop when it reaches $\mathbf{r_0}$ and gives rise to a diverging wave accounted for by $G_0^*(\mathbf{r}|\mathbf{r_0})$. As in a TR cavity experiment~\cite{cassereau,rosny,carminati}, the superimposition of the converging and diverging waves limits the focal spot size of the NR wave-field to one half wavelength [see Fig.~\ref{fig2}(c)]. The pre-factor $\rho/(1-\rho)$ in Eq.~(\ref{NR}) accounts for the multiple NR reflection events in the cavity [see Supplementary Information S2].

Depending on the sign of $\rho$, negative reflection can give rise to opposite interference phenomena. A reflection coefficient $\rho$ close to $-1$, as exhibited by the $S_1$ and $S_{2b}$ Lamb modes at a free edge~\cite{gerardin2016}, yields the following expression for $G(\mathbf{r}|\mathbf{r_0})$:
\begin{equation}
\label{NR2}
\lim\limits_{\rho \rightarrow -1} G(\mathbf{r}|\mathbf{r_0})= \mbox{Re} \left \{ G_0(\mathbf{r}|\mathbf{r_0}) \right \}.
\end{equation} 
The destructive interference between the incident and NR waves gives rise to a stationary wave-field that coincides with the real part of the free-space Green's function. In 2D or 3D configurations, this implies a singularity of the wave-field at the source location and a super-focusing ability [see Supplementary Information S2].

On the contrary, a positive sign of the reflection coefficient yields a stationary wave-field proportional to the imaginary part of the free-space Green's function:
\begin{equation}
\label{NR3}
\lim\limits_{\rho \rightarrow 1} G(\mathbf{r}|\mathbf{r_0}) = \frac{1}{\epsilon} \mbox{Im} \left \{ G_0(\mathbf{r}|\mathbf{r_0}) \right \},
\end{equation} 
with $\rho=1-\epsilon$ and $\epsilon <<1$. The constructive interference between the incident and NR waves gives rise to a resonant amplification of the wave-field accounted by the factor $\epsilon^{-1}$ in the last equation. As in a TR experiment~\cite{rosny}, the focal spot at the initial source location is diffraction-limited since $\mbox{Im} \left \{ G_0(\mathbf{r}|\mathbf{r_0}) \right \}$ exhibits a typical width of $\lambda/2$ [see Supplementary Information S2].

These fundamental results of Eqs.~(\ref{NR2}) and (\ref{NR3}) can also be extended to a source embedded into a random scattering medium at a depth much larger than the scattering mean free path. NR would induce a stationary wave-field and no energy could escape from the near-field of the source. The randomly disordered slab can therefore act as a band gap material. The observation of such super-focusing phenomena would require an experiment with an extremely fine frequency tuning~\cite{veres2016broad}. Moreover, it should also involve a non-invasive point-like source capable of a selective excitation of the forward or backward mode. Such experimental conditions are not met by the current experimental set up that operates in the time domain for a rigorous proof of the NR concept. An experimental demonstration of the NR super-focusing capabilities in the monochromatic regime thus constitutes a challenging perspective for this work.

From a more practical point-of-view, the experiment displayed in Fig.~\ref{fig4} illustrates how an ensemble of scatterers can become a NR mirror that retro-reflects the incident wave field towards the source itself. NR can thus be extremely useful for non-destructive testing applications. Lamb waves are indeed widely used in structural health monitoring for the detection of cracks in thin sheet materials and tubular products, such as aircraft shells or pipes for instance \cite{Boller,Su}. Usually, an image of the structure is built by means of a pulse echo method with a single transducer. However, a large part of the scattered wave-field is not collected by this transducer because of diffraction. This effect can severely limit the penetration depth of ultrasound imaging techniques because the echoes of interest are drowned into a predominant structure-borne noise. On the contrary, NR can backscatter all the energy of the scattered wave towards the emitting transducer and the signal of interest can then emerge from noise. As a consequence, NR can be taken advantage of for overcoming the imaging depth limit of standard pulse echo methods in structural health monitoring.

In summary, we have demonstrated the NR phenomenon in the time domain with elastic guided waves in plates of arbitrary shape. Whatever the {chaotic or random feature} of the plate, NR at its boundaries {or on the scatterers} gives rise to a replica of the incident wave, following the same path but in a reverse way. NR displays focusing abilities similar to a PC mirror while being a fully passive process. On the one hand, NR of elastic waves can be extremely useful in structural health monitoring but may also find applications in the development of new acoustic devices including resonators, lenses, and filters. On the other hand, {this phenomenon can also occur in zero-gap systems} such as photonic crystals~\cite{Sivan}, chiral metamaterials~\cite{Zhang} or graphene~\cite{Cheianov}. NR can thus become a tool of choice for wave-front shaping and focusing in all fields of wave physics. 

\section*{Methods}
\textbf{Experimental procedure.} The experiment consists in measuring the wave-field induced by a piezo-electric transducer or a piezo-electric strip glued on top of a plate [see Fig.\ref{fig1}(a) and (b)]. In the cavity, a $10~\mu$s chirp signal spanning the frequency range $1.8 - 2.0$~MHz is sent to the transducer, which generates a cylindrical incident wave front in the plate. In the disordered slab, a $15~\mu$s chirp signal spanning the frequency range $1.3 - 1.5$~MHz is sent to the piezoelectric strip, which generates a plane incident wave front. The out-of-plane component of the local vibration is detected on the other side of the plate by a photorefractive interferometer (Bossa Nova Tempo 1D). This probe is sensitive to the phase shift along the path of the optical probe beam. Signals detected by the optical probe are fed into a digital sampling oscilloscope and transferred to a computer. The corresponding wave field is measured over a time length of 1 ms and over a grid of points that map the plate surface with a pitch of 1 mm. 

\section*{Acknowledgements}

The authors are grateful for funding provided by LABEX WIFI (Laboratory of Excellence within the French Program Investments for the Future, ANR-10-LABX-24 and ANR-10-IDEX-0001-02 PSL*) and by the Agence Nationale de la Recherche (ANR-15-CE24-0014-01, Research Project COPPOLA). B.G. acknowledges financial support from the French ``Direction Générale de l'Armement'' (DGA).
\newpage

\renewcommand{\thefigure}{S\arabic{figure}}
\renewcommand{\theequation}{S\arabic{equation}}
\renewcommand{\thetable}{S\arabic{table}}
\setcounter{figure}{0} 

\begin{center}
\Large{\bf{Supplementary Material}}
\end{center}
\normalsize

This document provides further information on the separation of the incident and negatively reflected components of the wave-field recorded in the truncated billiard. It also provides the details of the calculations that lead to its theoretical expression in the monochromatic regime.

\section{Separation of the $S_{2b}$ and $S_1$ modes in the truncated billiard}

\begin{figure*}[!ht]
\centering
\includegraphics[width=14cm]{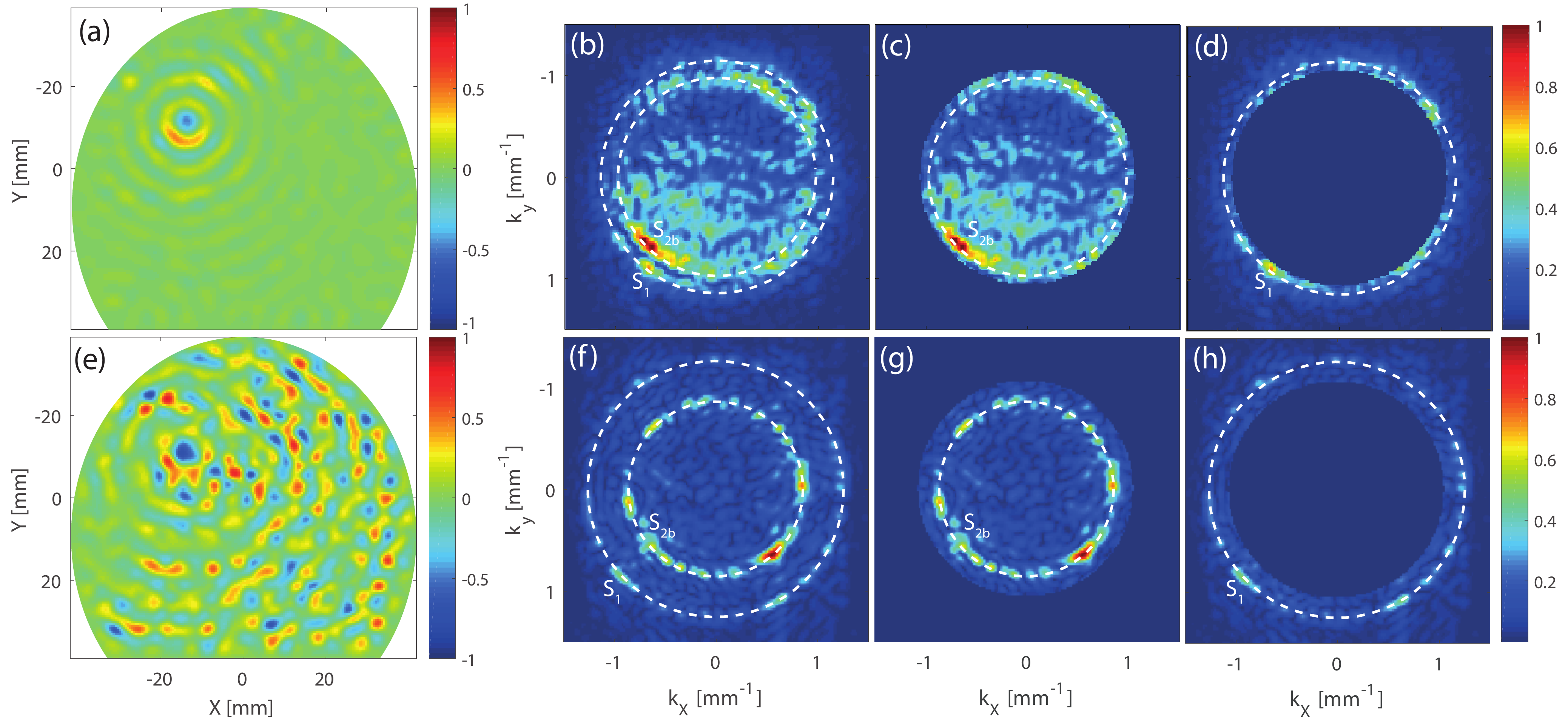}
\caption{Fourier analysis of the wave-field measured in the truncated billiard at the ZGV frequency $f = 1.91 $MHz (a-d) and above ZGV at $f=1.92$ MHz (e-h). From left to right : the wave fields (a,e), their spatial Fourier transforms (b,f), applied low-pass filter, $|k|<k_{ZGV}$, in order to isolate the $S_{2b}$ contribution (c,g), applied high-pass filter, $|k|>k_{ZGV}$, to isolate the $S_1$ contribution (d,h). 
}
\label{figS1}
\end{figure*}
Fig.~\ref{figS1}(a) displays the wave-field recorded by the interferometric probe at frequency $f=1.91$ MHz in the truncated billiard [see Fig.~2(a) of the accompanying paper]. Its spatial Fourier transform is displayed in Fig.~\ref{figS1}(b). It exhibits two rings in the vicinity of $k_{ZGV}=1.06$ mm$^{-1}$. The inner ring is associated with the $S_{2b}$ mode ($k_{2b}=0.975$ mm$^{-1}$) and the outer ring with the $S_1$ mode ($k_{1}=1.145$ mm$^{-1}$). To separate both contributions, spatial low and high pass filters are applied with a cut-off at $k_{ZGV}$. Fig.~\ref{figS1}(c) displays the result of the low pass filter in the spatial Fourier domain ($|k|<k_{ZGV}$). An inverse discrete Fourier transform (DFT) then yields the incident $S_{2b}$ wave-field shown in Fig.~3(a) of the accompanying paper. Fig.~\ref{figS1}(d) displays the result of the high pass filter in the spatial Fourier domain ($|k|>k_{ZGV}$). An inverse DFT then yields the negatively reflected $S_{1}$ wave-field shown in Fig.~3(c) of the accompanying paper. Fig.~\ref{figS1}(e) displays the wave-field recorded by the interferometric probe at frequency $f=1.92$ MHz. Its spatial Fourier transform is displayed in Fig.~\ref{figS1}(f). The inner ring is associated with the $S_{2b}$-mode ($k_{2b}=0.86$ mm$^{-1}$) and the outer ring with the $S_1$-mode ($k_{1}=1.26$ mm$^{-1}$). To separate both contributions, spatial low- and high-pass filters are applied with a cut-off at $k_{ZGV}$. Fig.~\ref{figS1}(g) displays the result of the low-pass filter in the spatial Fourier domain ($|k|<k_{ZGV}$). Fig.~\ref{figS1}(h) displays the result of the high-pass filter in the spatial Fourier domain ($|k|>k_{ZGV}$). Its inverse DFT then yields the reflected $S_{1}$ wave-field shown in Fig.~3(b) of the accompanying paper.

\section{Theoretical expression of the wave-field induced by negative reflection in a cavity}

Let us assume a point source at $\mathbf{r_0}$ in a chaotic cavity [see Fig.2(a) of the accompanying paper]. We assume that this source selectively emits a divergent forward wave (\textit{i.e.} of positive phase velocity). This choice is arbitrary and the same final result would be obtained if a divergent backward incident wave was considered. The incident wave-field $\phi_0(\mathbf{r})$ is given by the causal forward Green's function $G^{+}_{0}(\mathbf{r}|\mathbf{r_0})$ that would be obtained in free space. A first negative reflection event at the cavity boundary gives rise to a converging backward wave that mathematically corresponds to the anti-causal backward Green's function $G_{0}^{-*}(\mathbf{r}|\mathbf{r_0})$. As in a time-reversal experiment~\cite{rosny}, this wave collapses at the initial source point $\mathbf{r_0}$ and is always followed by a diverging wave. The first order reflected wave-field $\phi_1(\mathbf{r})$ thus shows two wave-fronts, the second one being the conjugate of the first one, but multiplied by $-1$:
\begin{equation}
\phi_1(\mathbf{r})= \rho \left [ G_{0}^{-*}(\mathbf{r}|\mathbf{r_0}) - G^-_{0}(\mathbf{r}|\mathbf{r_0}) \right ]
\end{equation}
where the symbol $*$ stands for phase conjugate and $\rho$ is the reflection coefficient. The diverging wave $-\rho G^-_{0}(\mathbf{r}|\mathbf{r_0})$ in the last equation is one more time negatively reflected. The second order reflected wave-field is then given by:
\begin{equation}
\phi_2(\mathbf{r})= \rho^2 \left [ G_{0}^+(\mathbf{r}|\mathbf{r_0}) - G^{+*}_{0}(\mathbf{r}|\mathbf{r_0}) \right ]
\end{equation}
The negative reflection process can be iterated. The reflected wave-field can be expressed after an even number of reflection events as,
\begin{equation}
\phi_{2n}(\mathbf{r})= \rho^{2n} \left [ G_0^{+}(\mathbf{r}|\mathbf{r_0}) - G_{0}^{+*}(\mathbf{r}|\mathbf{r_0}) \right ],
\end{equation} 
and after an odd number of reflection events as,
\begin{equation}
\phi_{2n+1}(\mathbf{r})= \rho^{2n+1} \left [ G_{0}^{-*}(\mathbf{r}|\mathbf{r_0}) - G_0^{-}(\mathbf{r}|\mathbf{r_0}) \right ]
\end{equation} 
By noting that $G_0^{-}(\mathbf{r}|\mathbf{r_0})=G_{0}^{+*}(\mathbf{r}|\mathbf{r_0})$ for strict negative reflection, the total wave-field $\phi(\mathbf{r})$ in the cavity can be finally expressed as
\begin{equation}
\label{aNR}
\phi(\mathbf{r})= \sum_{i=0}^{\infty}\phi_i(\mathbf{r})= G_{0}^+(\mathbf{r}|\mathbf{r_0}) + \frac{\rho}{1-\rho}\left [ G_{0}^+(\mathbf{r}|\mathbf{r_0}) - G_{0}^{+*}(\mathbf{r}|\mathbf{r_0}) \right ]
\end{equation} 

Depending on the sign of $\rho$, negative reflection can give rise to opposite interference phenomena. A reflection coefficient $\rho$ close to $-1$ yields the following expression for $\phi(\mathbf{r})$:
\begin{equation}
\label{aNR2}
\lim\limits_{\rho \rightarrow -1} \phi(\mathbf{r}) = \mbox{Re} \left \{ G_0(\mathbf{r}|\mathbf{r_0}) \right \}.
\end{equation} 
The destructive interference of the incident and NR waves gives rise to a stationary wave-field that coincides with the real part of the free-space Green's function. In 3D or 2D configurations, this implies a singularity of the wave-field at the source location and a super-focusing ability, since\cite{Watanabe}
\begin{equation}
\label{aNR4}
\mbox{Re} \left \{  G^{(3D)}_0(\mathbf{r}|\mathbf{r_0}) \right \} =\frac{\cos \left ( {2\pi}\|\mathbf{r}-\mathbf{r_0}\| / {\lambda}\right ) }{4 \pi \|\mathbf{r}-\mathbf{r_0}\|},
\end{equation} 
and 
\begin{eqnarray}
\label{aNR3}
\mbox{Re} \left \{  G^{(2D)}_0(\mathbf{r}|\mathbf{r_0}) \right \} &=& \frac{Y_0(2 \pi \|\mathbf{r}-\mathbf{r_0}\|/\lambda)}{4} \\ 
&\underset{\|\mathbf{r}-\mathbf{r_0}\|<<\lambda}{\sim} & \frac{1}{2 \pi} \left [\ln \left ( \frac{\pi \|\mathbf{r}-\mathbf{r_0}\|}{\lambda} \right ) + \gamma \right ] \nonumber
\end{eqnarray} 
with $Y_0$ the zero-order Bessel function of the second kind and $\gamma$ the Euler-Mascheroni constant~\cite{weisstein}.

On the contrary, a reflection coefficient $\rho$ close to $1$ yields the following expression for $\phi(\mathbf{r})$ [Eq.~(\ref{aNR})]:
\begin{equation}
\label{bNR3}
\lim\limits_{\rho \rightarrow 1} \phi(\mathbf{r}) = \frac{1}{\epsilon} \mbox{Im} \left \{ G_0(\mathbf{r}|\mathbf{r_0}) \right \},
\end{equation} 
with $\rho=1-\epsilon$ and $\epsilon << 1$. The constructive interference of the incident and NR waves gives rise to a resonant amplification of the wave-field at the source location accounted by the factor $\epsilon^{-1}$ in the last equation. As in a time reversal experiment~\cite{rosny}, the induced wave-field is given by the imaginary part of the free-space Green's function, which reads, in 2D and 3D, as follows
\begin{equation}
\label{bNR5}
\mbox{Im} \left \{  G^{(2D)}_0(\mathbf{r}|\mathbf{r_0}) \right \} = \frac{J_0(2 \pi \|\mathbf{r}-\mathbf{r_0}\|/\lambda)}{4} 
\end{equation} 
and 
\begin{equation}
\label{bNR6}
\mbox{Im} \left \{  G^{(3D)}_0(\mathbf{r}|\mathbf{r_0}) \right \} =\frac{\sin \left ( {2\pi}\|\mathbf{r}-\mathbf{r_0}\| / {\lambda}\right ) }{4 \pi \|\mathbf{r}-\mathbf{r_0}\|}.
\end{equation} 
Unlike the real part of $G_0(\mathbf{r}|\mathbf{r_0})$ [Eqs.~(\ref{aNR4})-(\ref{aNR3})], its imaginary part does not exhibit any singularity but shows a typical width of the order of $\lambda/2$. For $\rho\sim 1$, negative reflection thus gives rise to a diffraction-limited focal spot at the initial source location.

\end{document}